\documentclass[aps,10pt,prl,twocolumn,showpacs,groupedaddress]{revtex4-1}  
\usepackage{graphicx}  
\usepackage{dcolumn}   
\usepackage{bm}        
\usepackage{amssymb}   
\usepackage{amsmath}

\begin{document}

\title{Coupling between internal spin dynamics and external degrees of freedom in the presence of colored noise}
\author{S. Machluf}
	\email{machlufs@bgu.ac.il}
\author{J. Coslovsky}
\author{P. G. Petrov}
	\altaffiliation{Present address: Center for Cold Matter, Imperial College London, UK}
\author{Y. Japha}
\author{R. Folman}
	\affiliation{Department of Physics, Ben-Gurion University, Be'er Sheva 84105, Israel}
\date{\today}

\begin{abstract}
We observe asymmetric transition rates between Zeeman levels (spin-flips) of magnetically trapped atoms. 
The asymmetry strongly depends on the spectral shape of an applied noise. 
This effect follows from the interplay between the internal states of the atoms and their external degrees of freedom, where different trapped levels experience different potentials.
Such insight may prove useful for controlling atomic states by the introduction of noise, as well as provide a better understanding of the effect of noise on the coherent operation of quantum systems.
\end{abstract}

\maketitle

Coupling between internal and external (motional) degrees of freedom plays a major role in cooling atoms, ions and molecules~\cite{Dalibard_CCTannoudji:polarization_gradient, sideband, cooling_molecules} and in manipulating their quantum states, e.g., in logic gates for quantum computation \cite{quantum_gate1, quantum_gate, quantum_gate2, quantum_gate3}.
The state of these particles is usually controlled by monochromatic (or transform limited) electromagnetic fields, while incoherent fluctuations (noise) must be suppressed in order to prevent decoherence, heating  and loss~\cite{henkel:loss_and_heating, Jones:coupling_BEC_and_thermal_fluctuations, Vuletic:CP}. 
The latter are usually studied under the ``white noise'' assumption. Little is known about what happens between the monochromatic and white noise limits (``colored noise'') with respect to control and hindering effects.

To understand how the noise spectrum can affect the rate of transitions between internal states, consider a system of two levels representing electronic configurations of an atom, ion or molecule, coupled to external (translational, rotational or vibrational) degrees of freedom and to a weak homogeneous field inducing transitions between the internal levels. 
If this field imposes a monochromatic perturbation $\hat{\lambda}e^{-i\omega t}$, then the rate of transitions from an initial state $|i\rangle$ to a final state $|f\rangle$ is given by Fermi's golden rule $\gamma_{i\to f}= (2\pi/\hbar)\sum_k |\langle i| \hat{\lambda} |f,k \rangle |^2 \delta (E_i-E_{f,k} + \hbar\omega)$, where $k$ represents the quantum numbers of the external degrees of freedom of $|f\rangle$. 
The transition rates are proportional to the density of states in their respective final level, and are therefore asymmetric between two levels that experience different external potentials.
At the other extreme, one has a fluctuating random field $\hat{\lambda}g(t)$, where $g(t)$ has a spectral density $S(\omega)\equiv \int dt' e^{i\omega(t-t')}\langle g(t)g(t')\rangle$ which is flat over a large bandwidth (``white noise'').
This now allows transitions to all $k$ states whose completeness yields $\gamma_{i\to f}=S | \langle i| \hat{\lambda}|f \rangle|^2$ such that the external degrees of freedom decouple from the transition dynamics and the transition rates between the two levels become symmetric.

Numerous quantum systems proposed for applications such as quantum information processing are based on particles trapped in external potentials. 
A major limitation of these systems is uncontrolled noise-induced transitions between the internal states used for the application (e.g. qubits) or from the latter to other internal states.
Typically, different states may experience different potentials, and at times the potential difference is made significant as part of the application itself (e.g. quantum logic gates). 
As conventional environments usually contain background or technical noise which is not white, it is important to understand the interplay between the internal and external degrees of freedom under these conditions.

\begin{figure}[b]%
\includegraphics[width=\columnwidth]{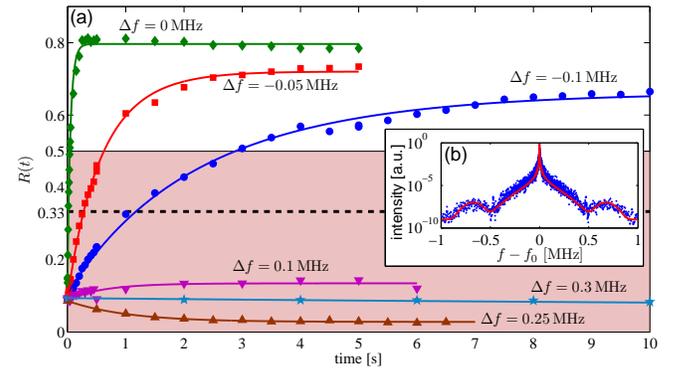}%
\caption{
(a) The measured ratio $R(t)$ between the number of atoms in $m_F=1$ and the number of atoms in both trapped levels. 
Different curves correspond to different values of $\Delta f$, the detuning between the noise peak frequency $f_0$ and the Zeeman splitting $E^0_{12}/h$ between the two trapped levels at the magnetic field minimum. 
The measured $R(t)$ goes beyond the band $0\leq R\leq 1/2$ representing asymptotic values of $R(t)$ in a model where the external degrees of freedom are decoupled from the transition dynamics (the dashed line at $R=1/3$ is for white noise). 
Curves are fits to the empirical form of Eq.~\eqref{eq:Rt_1}. 
Data variance gives an error of $\pm 0.01$ (not shown). 
(b) The noise spectrum we introduce into the system relative to $f_0=18\,$MHz (red line is the fit used in our calculations).
}%
\label{fig:expData}%
\end{figure}

In this work we study the dynamics of transitions between two internal atomic states with different external potentials in the form of two magnetically trapped Zeeman levels ($|F,m_F \rangle = |2,2 \rangle$ and $|2,1 \rangle$) in the presence of colored noise which is neither monochromatic nor white.
We experimentally observe that the relative transition rates between the levels strongly depend on the spectral shape of the noise. 
While a clear understanding of this dependence is important in order to find effective ways to combat uncontrolled noise, we demonstrate that this dependence also allows steering the transitions in the desired direction by utilizing engineered
noise.

We start our experiment with $\sim 7\times 10^4$ $^{87}$Rb atoms evaporatively cooled to $\sim350\,$nK. 
This leaves most of the atoms in the $|2,2\rangle$ Zeeman level with a residual fraction of $\sim$9\% in $|2,1\rangle$. 
The magnetic trap is produced 250$\,\mu$m from the surface of an atom chip by a right-angle Z-shaped wire~\cite{HD:UWire} and a homogeneous magnetic field generated by distant coils. 
This creates a harmonic trap with an axial frequency $\omega_x=2\pi\times 10\,$Hz (calculated) and radial frequencies $\omega_y=\omega_z=2\pi\times 96\,$Hz (measured) for the $|2,2\rangle$ level (a factor of $1/\sqrt{2}$ less for the $|2,1\rangle$ level). 
In order to study the effect of the noise spectrum we introduce controlled magnetic noise with the same function generator (Agilent 33250A) and antenna used for the evaporative cooling. 
To suppress uncontrolled noise, we increase the Zeeman splitting to 18MHz. 
At this range of spin-flip transition frequencies the uncontrolled noise spectrum is quite small and flat.
The populations $N_1$ and $N_2$ of the trapped levels $m_F=1$ and $m_F=2$, respectively,
are determined by a Stern-Gerlach measurement~\cite{ketterle:noise}.

Fig.~\ref{fig:expData}a presents the ratio $R(t)=N_1/(N_1+N_2)$ as a function of the time for which the noise is applied, for different detuning $\Delta f$ between the noise peak frequency $f_0$ and the Zeeman splitting $E^0_{12}/h$ of the two trapped levels at the magnetic field minimum.
Fig.~\ref{fig:expData}b presents the spectrum of the noise we introduce. 
One may observe a large difference in the evolution of $R(t)$ between red- and blue-detuned noise. When the noise is red-detuned ($\Delta f < 0$), the $m_F=1$ level is populated while when the noise is blue-detuned ($\Delta f > 0$) the $m_F=1$ level is depleted. 
The solid curves are fits to the empirical form
\begin{equation}
	R(t)=R_{\infty}+(R_0-R_{\infty})e^{-\tilde \gamma t},
\label{eq:Rt_1}
\end{equation}
which represents an exponential convergence from the initial value $R_0 \equiv R(t=0)$ to an asymptotic value $R_{\infty} \equiv R(t \to \infty)$. These results show that the relative transition rates between the levels strongly depend on the detuning of the noise and significantly differ from the expected rates in the case of white noise (dashed line in Fig.~\ref{fig:expData}a). The value of $\tilde \gamma$ in Eq.~(\ref{eq:Rt_1}) strongly depends on the intensity of the noise at the relevant frequency.
At resonance ($\Delta f=0$) it is roughly hundreds of Hz, depending on the exact noise amplitude used. 

The transitions between the atomic levels are caused by the coupling $\hat{\lambda}= -g_F\mu_B\hat{{\bf F}}\cdot {\bf B}_{\rm noise}$ between the magnetic field noise ${\bf B}_{\rm noise}$ and the magnetic  moment of the atom. 
Here $\mu_B$ is the Bohr magneton, $\hat{{\bf F}}$ is the angular momentum operator and  $g_F$ is the Land\'e factor. 
This coupling allows only transitions with spin change $|\Delta m_F|=1$, which implies the following rate equations for the populations of the two trapped levels
\begin{eqnarray}
	\dot{N}_1 &=& -(\gamma_{1 \to 2}+\gamma_{1\to 0})N_1+\gamma_{2\to 1}N_2 \nonumber \\
	\dot{N}_2 &=& \gamma_{1\to 2}N_1-\gamma_{2\to 1}N_2.
\label{eq:rateEq}
\end{eqnarray}
We estimate the escape time of an atom in the untrapped level $m_F=0$ from the trapping region, due to thermal velocity and gravitational acceleration, to be $\sim 10\,$ms. 
For $|\Delta f|>20\,$kHz ($\pm 10\,$kHz due to the variance in $\tilde \gamma$) the atom is lost from the trap before it can make a transition back to $m_F=1$ and the transition $\gamma_{0\to 1}$ may be safely neglected. 
In the limit of white noise, where the motional and internal degrees of freedom are decoupled,
we expect the ratio between the transition rates to be determined by the matrix elements of the angular momentum operator, such that $\alpha\equiv \gamma_{1\to 0}/\gamma_{2\to 1}=3/2$, while $\beta\equiv \gamma_{1\to 2}/\gamma_{2\to 1}=1$~\cite{Vuletic:CP, valery:reduction_mag_noise}. In the following we show that colored noise dictates different values for $\alpha$ and $\beta$.

\begin{figure}%
\includegraphics[width=\columnwidth]{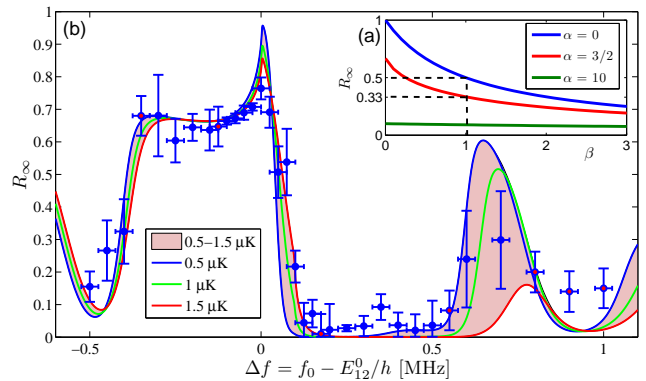}%
\caption{
(a) The dependence of $R_\infty$ [Eq. \eqref{eq:RInf}] on $\beta$ for various values of $\alpha$. (b) Asymptotic values $R_{\infty}$ taken from experimental curves as in Fig. \ref{fig:expData}, as a function of $\Delta f$, compared with theory [Eqs.~\eqref{eq:RInf}-\eqref{eq:gamma1}] with no fitting parameters ($T=0.5-1.5\,\mu$ K).
Error bars are rms of data variance and mean fit error in Fig.~\ref{fig:expData}, except for points marked in red, for which only one measurement is available and the error is the average of the whole data set. 
The horizontal error is the uncertainty in the magnetic field minimum.
}%
\label{fig:fInf}%
\end{figure}

Next we connect the ratio $R(t)$ to $\alpha$ and $\beta$ and focus on the asymptotic value $R_\infty$.
Under the assumption that the transition rates are time-independent, Eqs.~\eqref{eq:rateEq} yield the solution
\begin{equation}
	R(t)\equiv \frac {N_1(t)}{N_1(t)+N_2(t)}=\frac{R_\infty+Ce^{-\tilde \gamma t}}
		{1+\alpha R_\infty C e^{-\tilde \gamma t}}
\label{eq:Rt}
\end{equation}
where $C\equiv(R_0-R_\infty)/(1-\alpha R_\infty R_0) \text{ and }\tilde \gamma\equiv (1/R_\infty-\alpha R_\infty)\gamma_{2\to 1}$. 
Note that Eq.~\eqref{eq:Rt} reduces to Eq.~\eqref{eq:Rt_1} when $\alpha=0$, and even when $\alpha \gg 1$ Eq.~\eqref{eq:Rt_1} is a fair approximation. The asymptotic value in Eq.~\eqref{eq:Rt} is given by
\begin{equation}
	R_\infty = \frac {1+\alpha+\beta - \sqrt{-4\alpha+(1+\alpha+\beta)^2}}{2\alpha}.
\label{eq:RInf}
\end{equation}
This value depends solely on the ratios $\alpha$ (determining the relative depletion rate to the $m_F=0$ level) and $\beta$ (determining the asymmetry in the transition rate between the two trapped levels), and is independent of the overall noise intensity or the initial condition, $R_0$. 
In the white noise limit, where $\alpha=3/2$ and $\beta=1$, $R_{\infty}=1/3$ (dashed line in Fig.~\ref{fig:expData}a) and $R_{\infty}\leq 1/2$ for any value of $\alpha$ ($\beta=1$ band in Fig.~\ref{fig:expData}a). 
The dependence of Eq.~\eqref{eq:RInf} on $\beta$ for various values of $\alpha$ is presented in Fig. \ref{fig:fInf}a.

In Fig.~\ref{fig:fInf}b we plot the measured value of $R_\infty$ for different values of the center frequency of the applied noise. 
The band and the three curves represent the theoretical prediction for $R_\infty$, as described below.

\begin{figure}%
\includegraphics[width=\columnwidth]{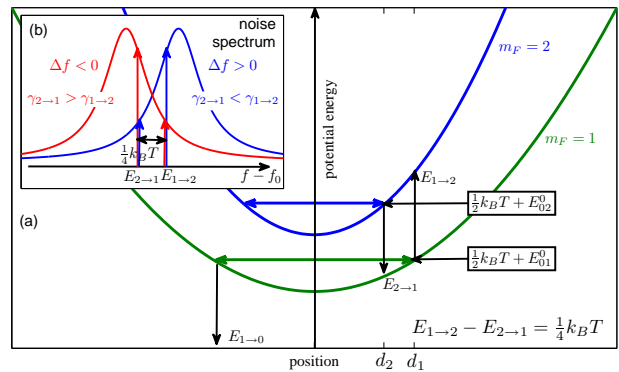}%
\caption{
(a) A simplified model showing the asymmetry in the transitions between the two trapped levels. 
We assume that the atoms are in thermal equilibrium at temperature T, in both levels. 
We also assume momentum to be unchanged during the transition as the photon recoil is negligible, and this is why the length of the arrows corresponds to the difference between the potentials. 
We plot the transitions at the mean atom distances from the trap center, $d_j\equiv\sqrt{\langle x^2\rangle_j}$. 
A spin flip from $m_F=2$ to $m_F=1$ requires photon energy $E_{2\rightarrow 1}$, which is smaller than the photon energy $E_{1\rightarrow 2}$ needed for the reverse transition. 
(b) The two transitions sample different frequencies of the noise spectrum (plotted twice for opposite detuning), giving rise to the asymmetric  transition rates. 
The blue and red pairs of arrows point to the sampled noise intensities for each detuning.
}
\label{fig:explanation}%
\end{figure}

In order to gain a qualitative understanding of the results, we first present a simple semiclassical 1D model of two thermal distributions of atoms trapped in potentials $V_j(x) = \frac {1} {2} m_j M \omega_1^2 x^2$, where $M$ is the atomic mass, $\omega_1$ is the trapping frequency of the $m_F=1$ level and $m_j=m_F$ is either 1 or 2. 
Each atomic distribution is represented in Fig.~\ref{fig:explanation} by a typical atom at an average position $d_j=\sqrt{\langle x^2\rangle_j}=\sqrt{k_B T/m_j M\omega_1^2}$. 
As the wavelength of the applied noise is much larger than the typical size of the system, we may assume that momentum is conserved locally during the transition (the recoil is negligible).
The transition of a typical atom from $m_F=2$ to $m_F=1$ requires a photon energy $E_{2\to 1}=E^0_{12}+ V_2(d_2)-V_1(d_2)
=E^0_{12}+\frac{1}{4}k_B T$ and reduces the energy of the atom relative to the trap bottom by $\frac{1}{4}k_B T$.
A transition of a typical atom in $m_F=1$ to $m_F=2$ at $d_1=\sqrt{2}d_2$ requires a photon energy 
$E_{1\to 2}=E^0_{12}+ V_2(d_1)-V_1(d_1)=E^0_{12}+\frac 1 2 k_B T$  and increases the energy of the atom  by $\frac{1}{2}k_B T$. 

The relative transition rates depend on the number of photons with the two energies
$E_{1\to 2}> E_{2\to 1}$ .
If the noise intensity increases with frequency in a wide enough band ($\sim k_B T$) around $f=E_{12}^0/h$ (see blue arrows in Fig.~\ref{fig:explanation}b, $\Delta f>0$), the transition $1\to 2$ is preferred and the population of $m_F=1$ is depleted, as in the plateau on the right hand side of Fig.~\ref{fig:fInf}b.
If, by contrast, the intensity decreases in this band (see red arrows in Fig.~\ref{fig:explanation}b, $\Delta f<0$), the transition $2\to 1$ is preferred and the population of $m_F=1$ becomes dominant, as in the plateau on the left-hand side. 
Due to these effects, even when the noise intensity decreases by a few orders of magnitude (side peaks of the noise), the atoms are still sensitive to the details of the spectrum, as can be seen in Fig. \ref{fig:fInf}b. 
This may prove to be a useful tool for characterizing noise features. 
Note also that the temperature determines the width of the spectral region the cloud samples, hence colder clouds are more sensitive to the fine details of the noise (e.g. 1$\,\mu$K gives a resolution of 20$\,$kHz). 
This effect is most noticeable at $\Delta f \approx 0.7\,$MHz (Fig. \ref{fig:fInf}b) where the temperature band becomes much wider.

To quantitatively explain the results, the ratios $\alpha$ and $\beta$ are calculated using a semiclassical expression for the transition rate which follows from Fermi's golden rule \cite{japha:gamma}
\begin{equation}
	\gamma_{i\to f} = \int d\omega \Gamma_{if}(\omega)\int \frac {d^3{\bf p} d^3{\bf r}} {(2\pi\hbar)^{3}}  P_i({\bf r},{\bf p})
	\delta (\Delta V_{if}({\bf r}) - \hbar \omega)
\label{eq:gamma}
\end{equation}
where $\Gamma_{if}(\omega) = g_F^2 \mu_B^2 \sum_{j=y,z} |\langle i| \hat{F}_j|f\rangle|^2 S_B^{jj}(\omega)$, $S_B^{jj}$ being the spectral density of the magnetic fluctuations~\cite{valery:reduction_mag_noise, henkel:loss_and_heating}, and $P_i({\bf r}, {\bf p})$ is a normalized phase space distribution for level $i$.
$\Delta V_{if} = E_{if}^0+V_f-V_i$, where $E_{if}^0$ is the Zeeman splitting calculated using the Breit-Rabi formula, predicting a difference of $h \times 95\,$kHz between $E^0_{12}$ and $E_{01}^0$. $V_j = \frac 1 2 m_j M \sum_{k=x,y,z} \omega^2_{1k} r_k^2 +Mgz$, where $\omega_{1k}$ is the trapping frequency of the $m_F=1$ level along the $k$th axis (frequencies given above), $m_j\ (j=i,f)$ are the indices of the initial and final Zeeman levels, and $g$ is the gravitational acceleration which causes a shift of the trap centers relative to each other.
Assuming a Maxwell-Boltzmann distribution, $P_i\propto e^{-p^2/2mk_B T}e^{-V_i/k_B T}$ with a constant temperature $T$ for both levels, Eq.~\eqref{eq:gamma} reduces to an integral over dimensionless space coordinates
\begin{eqnarray}
	\gamma_{i \to f} &=& 4\frac{m_i^{3/2}}{\sqrt{\pi}}\int_0^{\infty}{dqq^2 e^{-(m_iq^2+\eta^2/m_i)} }
			\frac {\sinh(2\eta q)}{2\eta q} \times \nonumber \\
	&\times&  \Gamma_{if}(\omega=(E_{if}^0+q^2k_B T)/\hbar),
\label{eq:gamma1}
\end{eqnarray}
where $\eta = (\text{g}/\omega_{1z})\sqrt{M/2k_B T}$. 
Trap frequency variations due to the non-linear Zeeman effect are neglected.

The theoretical curves are calculated utilizing a fit to the independently measured spectrum of the noise (red curve in Fig.~\ref{fig:expData}b): a Lorentzian of $1\,$kHz FWHM  multiplied by a Gaussian with $\sigma=150\,$kHz for the center peak, two identical Gaussians for the two side peaks, and a constant for the background white noise. 
Finer details of the noise are ignored.
The band in Fig.~\ref{fig:fInf}b represents the different theoretical values of $R_\infty$ for cloud temperatures ranging from 0.5$\,\mu$K to 1.5$\,\mu$K. 
The $1\,\mu$K wide band accounts for heating due to the transition process
($\frac{1}{4}k_B T$ per $2\to 1\to 2$ cycle in Fig.~3) and due to uncontrolled background noise.

\begin{figure}
\includegraphics[width=\columnwidth]{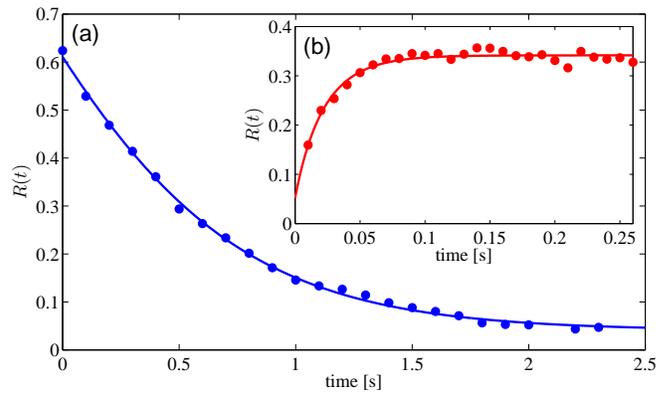}%
\caption{
(a) Demonstration of control by noise. 
We use red-detuned noise ($\Delta f=-0.2\,$MHz) for 200$\,$ms transferring most of the atoms into level $m_F=1$ (as predicted in Fig.~\ref{fig:fInf}b). 
We then ($t\equiv 0$) apply blue-detuned noise ($\Delta f=+0.4\,$MHz) which transfers all the atoms into level $m_F=2$, as predicted by theory (Fig.~\ref{fig:fInf}b). 
(b) Measured ratio $R(t)$ for applied white noise. 
The measured asymptotic value of $R_{\infty}=0.344\pm 0.004$ is very close to the predicted $R_{\infty}=1/3$. 
Errors are estimated as in Fig.~\ref{fig:expData}.
}%
\label{fig:jumpWhiteNoise}%
\end{figure}

To further verify our model and to demonstrate control by noise, we conducted the following experiments: first, we introduce red-detuned noise ($\Delta f = -0.2\,$MHz). 
At this frequency the system has a steady state of $R_\infty \approx 0.7$, as can be seen in Fig.~\ref{fig:fInf}b. 
When most of the atoms are in the $m_F=1$ level, we rapidly change the center frequency to $\Delta f = +0.4\,$MHz. 
In the new steady state all atoms should be in the $m_F=2$ level, exactly as we have observed (Fig.~\ref{fig:jumpWhiteNoise}a).

It is interesting to consider our model under the two limits noted previously, namely, white noise and monochromatic radiation. 
For white noise our theory predicts $R_\infty=1/3$, which is confirmed by our measurement presented in Fig.~\ref{fig:jumpWhiteNoise}b.
The prediction of our model for the other limit may be estimated by introducing into Eq.~\eqref{eq:gamma1} $\Gamma_{if}(\omega) \propto \delta(\omega-\omega_{\rm{noise}})$.
It follows that transitions occur only for $\Delta f \geq 0$ and that the ratio $\beta$  becomes
\begin{equation}
	\beta_{\rm mono} = 2^{-3/2}\exp\left(\frac{2\pi\hbar\Delta f - Mg^2/4\omega_{1z}^2}{k_B T}\right).
\end{equation}
The factor $2^{-3/2}$ is the 3D ratio between the density of states in the two levels.
The second factor in the exponent is due to the different gravitational shift of the atomic potential minima.
When the transition $1\to 0$ is far detuned from the transition $2\to 1$ ($\alpha\to 0$) due to the
non-linear Zeeman shift, $\lim_{\alpha \to 0}R_{\infty} = 1/(1+\beta) \approx$ 0.92 (1$\,\mu$K), 0.85 (2$\,\mu$K), relative to a value of $R_{\infty}\approx$ 0.73 expected from the density of states ratio alone.
This explains the high theoretical values in Fig. \ref{fig:fInf}b near resonance, where the noise
is dominated by a sharp peak.
We attribute the fact that $R_\infty \leq 0.8$ in all our near-resonance measurements to the fast heating rates which were observed independently.
Note also that in this region our assumption regarding thermal equilibrium is invalid.
At $|\Delta f|<150\pm50\,$kHz, the collision rate (estimated to be $\sim 0.5\,$Hz) is slower than the spin flip transition rates estimated by $\tilde\gamma$.
Nevertheless, asymmetry exists in this regime as well, which may indicate that thermalization is not the fundamental source of the observed asymmetry.
Indeed, in Eq.~\eqref{eq:gamma1} we have used the Maxwell-Boltzmann distribution as a mere simplifying assumption and we believe that a more elaborate model would reach similar results without invoking thermal equilibrium.

To summarize, we have observed that the spectral shape of noise determines the relative transition rates between internal levels of trapped neutral atoms, due to the interplay of these levels with the external degrees of freedom. 
As non-white noise spectra are common in background and technical noise (e.g. in the radio-frequency regime), and may be very significant in some types of apparatuses such as atom chips \cite{RonRev}, these findings may serve to better understand how noise couples to quantum systems based on trapped atoms. 
Future studies should investigate the effects of colored noise on the coherence in systems of trapped particles with few levels, and the possibility that colored noise may be used for the suppression of decoherence, as proposed in Ref.~\cite{Kurizki}.  

RF thanks Yoseph (Joe) Imry for being a profound mentor. 
SM would like to thank Julien Chab\'{e} for helpful discussions. 
We thank A. Aharony, O. Entin-Wohlman, B. Horovitz and M. Schechter for stimulating discussions.
We acknowledge support from the German-Israeli fund (GIF) and from the Israeli Science foundation (ISF).

\bibliographystyle{apsrev4-1} 

%

\end{document}